\documentclass[11pt, a4paper]{article}

\usepackage[utf8]{inputenc}
\usepackage[T1]{fontenc}
\usepackage{mathptmx} % Times New Roman benzeri font
\usepackage{geometry}
\usepackage{amsmath}
\usepackage{xcolor}
\usepackage{graphicx}
\usepackage{indentfirst} % İlk paragraflara girinti eklemek için

\usepackage{natbib}
\usepackage{orcidlink}
\usepackage{hyperref}

\usepackage{sectsty}
\sectionfont{\centering}
\usepackage{setspace}
\usepackage[labelfont=bf]{caption}
\begin{document}

% Üst Bilgi (Header) Alanı
\noindent
\begin{tabular*}{\textwidth}{@{} l @{\extracolsep{\fill}} r @{}}
\textcolor{gray}{\small Aegean J. Theor. Comput. Appl. Sci. xx (xxxx) zzz-zzz} & \textcolor{gray}{\small Published by Faculty of Sciences} \\
\textcolor{gray}{\small } & \textcolor{gray}{\small Ege University, Türkiye} \\
& \textcolor{gray}{\small https://dergipark.org.tr/xxxxx}
\end{tabular*}

%\vspace{3em}

% Başlık Alanı
\begin{center}
{\Large \bfseries GEOMAGNETIC STORM IMPACTS ON THE IONOSPHERE \\[0.3em] OVER TÜRK\.{I}YE DURING 
SOLAR CYCLE 25: \\[0.3em] FOCUSING ON THE MAY 2024 
STORM \\[0.6em] \par} 

%\vspace{2em}

% Yazarlar
{\normalsize \bfseries 
E. Erayd{\i}n\orcidlink{0009-0007-8615-9600}$^1$, 
S. Ta\c{s}demir\orcidlink{0000-0003-1339-9148}$^{2,*}$, 
D. C. \c{C}{\i}nar\orcidlink{0000-0001-7940-3731}$^2$, 
S. Öz{\i}rmak\orcidlink{0009-0006-3854-5286}$^2$ and 
R. Canbay\orcidlink{0000-0003-2575-9892}$^3$ \\[0.6em] \par}

%\vspace{1.5em}

{\itshape \small
$^1$ Yeditepe University, Faculty of Arts and Sciences, Programme of Physics, 34640, Ata\c{s}ehir, Istanbul, Türkiye \\[0.5em]
$^2$ Istanbul University, Institute of Graduate Studies in Science, Programme of  Astronomy and Space Sciences, \\ 34116, Beyaz{\i}t, Istanbul, Türkiye \\[0.5em]
$^3$ Akdeniz University, Faculty of Science, Department of Space Sciences and Technologies, 07058, \\ Antalya, Türkiye \par}
\end{center}
\date{XX XX, 2025}

%\maketitle

\noindent {\bfseries Abstract:} The interaction between solar activity and Earth's magnetosphere magnetosphere-ionosphere system often results in geomagnetic storms that disturb ionospheric electron density. In this study, we analyse the ionospheric response to selected geomagnetic storm events during the Solar Cycle 25, focusing on the mid latitude region of Earth, including Turkey. Hourly Kp and Dst indices obtained from the OMNI database are compared with global TEC maps provided by NASA CDDIS. Storm time anomalies include short term enhancements and irregularities in electron content, correlated with geomagnetic activity. Unlike equatorial regions, mid-latitude ionospheric responses exhibit distinct features such as Storm Enhanced Density (SED). These findings emphasize the importance of continuous space weather monitoring for navigation and communication systems.

\vspace{1em}
\noindent \textbf{Keywords:} Solar Activity -- Solar Observations -- Solar Cycle 25
%\end{abstract}

\section{Introduction}

Solar activity, through phenomena such as solar flares and coronal mass ejections (CMEs), constitutes the primary driver of space weather variability in the heliosphere. The interaction between solar transients and the Earth's magnetosphere ionosphere system generates geomagnetic storms and global disturbances that perturb plasma density, electric fields, and current systems across multiple geo-space regions \citep{Gopalswamy2022, Temmer2021}. These storms, resulting from enhanced solar wind magnetosphere coupling, cause significant fluctuations in ionospheric Total Electron Content (TEC), thereby impacting satellite navigation, communication, and positioning systems \citep{Bagheri2025, Shahzad2023}.

Recent case studies based on GNSS derived TEC data have revealed latitude dependent storm time behaviours. In the equatorial sector, \citet{Lopez2025} analysed the G5 geomagnetic superstorm that occurred over Ecuador and the Galápagos Islands in May 2024 and reported a gradual recovery following an unusual TEC depletion. This situation was attributed to a model involving complex reconnection and plasma redistribution processes in the equatorial ionosphere. In contrast, mid-latitude regions typically exhibit storm time phenomena such as Storm Enhanced Density (SED) and traveling ionospheric disturbances (TIDs), which can significantly degrade navigation and communication performance \citep{FullerRowell2008}. These diverse responses emphasize the importance of continuous ionospheric monitoring across different latitude sectors.

In the mid-latitude sector, particularly over Türkiye, recent geomagnetic storms have revealed pronounced storm time depletions in TEC, followed by partial post storm recovery phases. The May 2024 G5 superstorm, one of the most intense events of Solar Cycle~25, produced TEC reductions from nearly 50 TECU to 15 TECU, indicating a strong coupling between the solar wind and the ionosphere in this region. Unlike the equatorial response reported by \citep{Lopez2025}, which exhibited an atypical depletion pattern associated with equatorial electrodynamics, the mid-latitude ionosphere over Türkiye displayed a more regular but stronger depletion signature. This highlights the latitude dependent nature of ionospheric storm time dynamics, underscoring the necessity of incorporating regional analyses into global space weather studies.

In this study, we investigate the mid-latitude ionospheric response over Türkiye during Solar Cycle~25 by analysing GNSS derived TEC maps from NASA~CDDIS in comparison with geomagnetic activity indices ($Kp$ and $Dst$) obtained from the OMNI database. The correlation between geomagnetic drivers and ionospheric electron content provides insights into storm time ionospheric variability at mid-latitudes.

\section{Data}

To investigate the ionospheric response to geomagnetic storms during the ascending phase of Solar Cycle~25, we employed a multi dataset approach integrating solar wind, geomagnetic, and ionospheric parameters over Türkiye region. The datasets used in this study are summarized below:

\textbf{OMNI Data:}  
Interplanetary parameters and geomagnetic indices were obtained from NASA’s OMNIWeb database\footnote{Data were accessed from the NASA/GSFC OMNIWeb service: \url{https://omniweb.gsfc.nasa.gov}}. The dataset provides hourly measurements of solar wind plasma properties (velocity $V_{sw}$, density $n_p$, dynamic pressure $P_{dyn}$), interplanetary magnetic field (IMF) components ($B_x$, $B_y$, $B_z$, and $|B|$), and key geomagnetic indices ($Kp$, $Dst$, $AE$, $AL$, $AU$, and $F_{10.7}$). 
These parameters serve as proxies for magnetospheric energy input and have been widely utilized in previous studies examining storm time dynamics \citep[e.g.,][]{Gonzalez1994, Blagoveshchenskii2013, Lakhina2016}.

\begin{figure}[h!]
\centering
\includegraphics[width=0.6\textwidth]{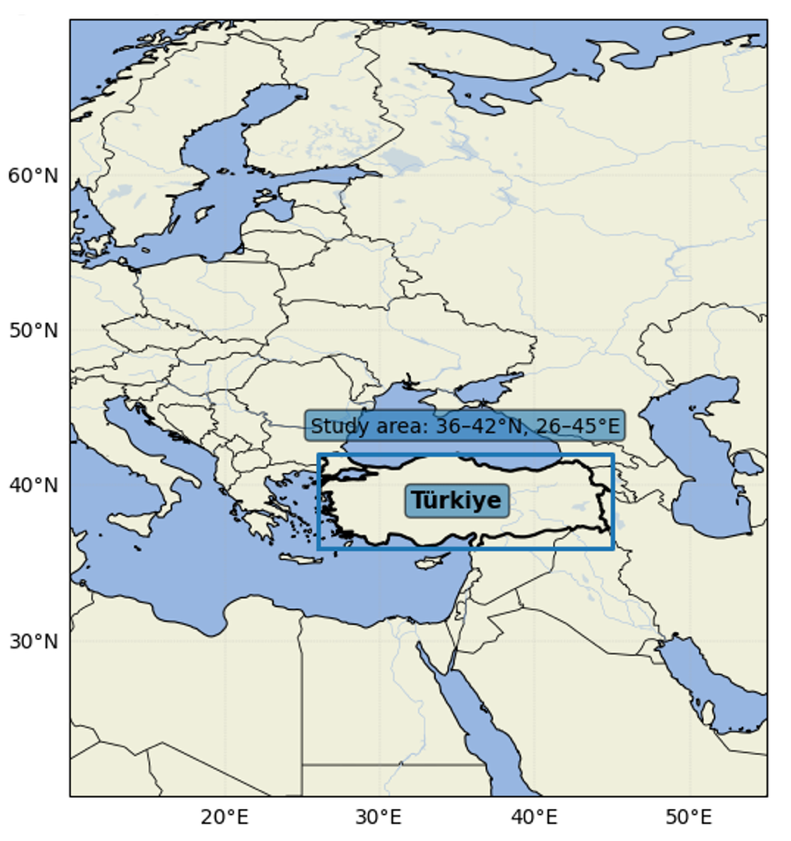}
\caption{
The area analyzed in this study, covering the geographical region of Türkiye between $36^{\circ}$–$42^{\circ}$~N latitude and $26^{\circ}$–$45^{\circ}$~E longitude.}
\label{fig:fig1}
\end{figure}

\textbf{GNSS based Ionospheric Data:}
Global Ionospheric Maps (GIMs) in IONEX format were retrieved from the NASA CDDIS archive\footnote{Global Ionospheric Maps (GIMs) were accessed from the NASA CDDIS archive: \url{https://cddis.nasa.gov/archive/gnss/products/ionex}}. These maps provide Vertical Total Electron Content (VTEC) values with a spatial resolution of $2.5^{\circ}\!\times\!5^{\circ}$ and a temporal resolution of 2~hours \citep{Li2023}. For regional analyses, we extracted the subset covering $36^{\circ}$–$42^{\circ}$~N and $26^{\circ}$–$45^{\circ}$~E encompassing Türkiye and adjacent mid-latitude zones and resampled it to an hourly cadence Figure~\ref{fig:fig1}. This approach follows the methodology applied by \citet{Lopez2025}, ensuring consistency with global TEC analyses.However, due to the coarse spatial resolution of GIMs, mesoscale and small-scale ionospheric structures such as storm-enhanced density features or travelling ionospheric disturbances-cannot be reliably resolved. Therefore, the results presented in this study primarily reflect large-scale ionospheric behavior over the region, and finer-scale variability may remain undetected in the current dataset.
To contextualize the May 2024 superstorm within Solar Cycle 25, we also inspected ionospheric conditions during a severe G4 event in April 2023 and a minor G1 storm in November 2022. Both events exhibited substantially weaker TEC perturbations compared to the May 2024 case, confirming that the selected superstorm represents an extreme upper-limit disturbance during the cycle, while typical mid-latitude ionospheric variability under lower-intensity storms remains much more moderate Table~\ref{tab:table1}.

\begin{table}[h!]
\centering
\caption{Event intensities are classified according to NOAA G-scales (G1--G5).}
\label{tab:table1}
\begin{tabular}{lcccc}
\hline
Date (UTC) & Dstmin (nT) & Kpmax & NOAA G-class \\
\hline
2024-05-11 & $-412$ & 9  & G5 \\
2023-04-24 & $-213$ & 8  & G4 \\
2023-11-05 & $-172$ & 7+ & G3 \\
2022-11-07 & $-92$  & 5  & G1 \\
\hline
\end{tabular}
\end{table}

\section{Methods}

Storm events were selected using thresholds of Dst~$<\!-200$~nT and Kp~$\geq\!8$, corresponding to severe to extreme geomagnetic activity (G4--G5) \citep{Gonzalez1994, NOAAGscale}. The primary case examined is the May 11, 2024, superstorm, which was triggered by multiple Earth-directed CMEs originating from NOAA Active Region 3664 and resulted in G5 conditions \citep{SWPCG520240511,SWPCG4Watch20240510}. To place this event in context, we also refer to a severe G4 storm from April 2023 and a minor G1 episode from November 2022 \citep{SWPCG420230423,SWPCG1202211}.

All data series were transformed to Universal Time (UT) and interpolated to a uniform hourly resolution. Outliers and data gaps were corrected using a rolling mean and linear interpolation technique. TEC data were normalized by subtracting the quiet day mean (typically 3–5 days before the storm) to emphasize storm time deviations \citep{Remya2025, Lopez2025}.
\begin{equation}
\Delta \mathrm{TEC}(t) = \mathrm{TEC}(t) - \overline{\mathrm{TEC}}_{\text{prestorm}}.
\end{equation}
This normalization enables comparison of relative ionospheric perturbations independent of background diurnal variations.

The interdependence between geomagnetic activity and ionospheric variability was quantified by computing Pearson correlation coefficients between the geomagnetic indices ($Kp$ or $Dst$) and the corresponding \(\Delta\)TEC series over a lag window of $\pm$12~hours:
\begin{equation}
r(\tau) = \frac{\sum_{t} [x(t) - \bar{x}][y(t+\tau) - \bar{y}]}{\sigma_x \sigma_y},
\end{equation}
where \(x(t)\) and \(y(t)\) denote the geomagnetic index and the \(\Delta\)TEC time series, respectively, and \(\tau\) represents the temporal lag. Negative lags indicate that ionospheric perturbations precede geomagnetic variations, potentially linked to prompt penetration electric fields or preconditioning effects. 

Conversely, positive lags correspond to delayed ionospheric responses during the main and recovery phases, primarily governed by thermospheric composition changes and neutral wind dynamics. A lag window of ±12~hours was selected to capture both the immediate and delayed coupling processes frequently reported during intense geomagnetic storms. Similar correlation based frameworks have been employed in previous mid and low latitude studies examining storm time coupling \citep[e.g.,][]{Cherniak2018, Yang2022}.

Each geomagnetic storm was divided into three canonical phases based on the temporal evolution of the $Dst$ and $Kp$ indices \citep{Gonzalez1994};

\begin{itemize}
    \item Initial phase: Sudden commencement (SSC) characterized by a rapid $Kp$ increase and the initial Dst depression.
    
    \item Main phase: Sharp $Dst$ decrease reaching minimum values.
    
    \item Recovery phase: Gradual return of $Dst$ toward quiet levels.TEC variations were analysed across these intervals to distinguish negative and positive storm effects, following the frameworks of \citet{Blagoveshchenskii2013} and \citet{Remya2025}.
\end{itemize}

\begin{figure}[h!]
\centering
\includegraphics[width=1\textwidth]{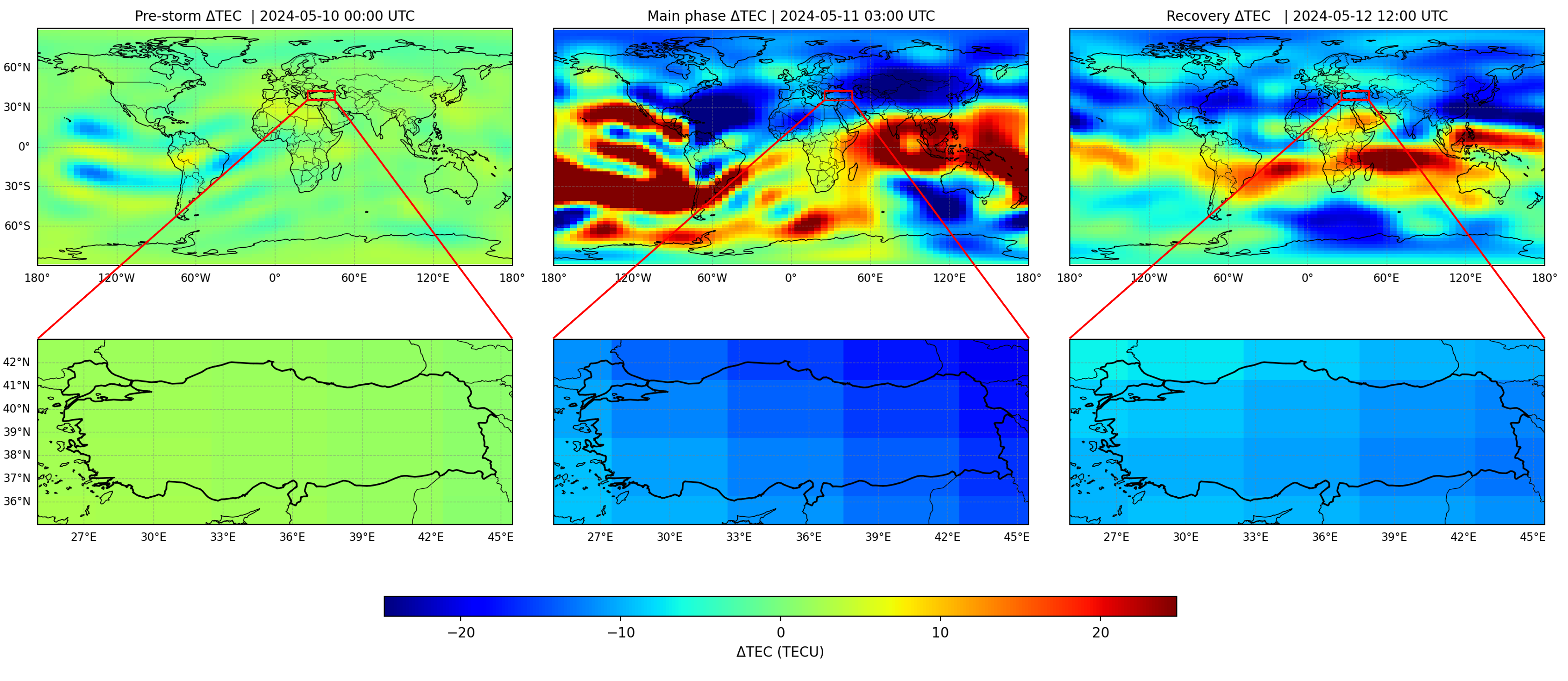}
\caption{Two-dimensional TEC anomaly ($\Delta \mathrm{TEC}$) maps for the pre-storm (2024-05-10 00:00~UTC), main phase (2024-05-11 03:00~UTC), and recovery phase (2024-05-12 12:00~UTC) of the geomagnetic storm. Global $\Delta \mathrm{TEC}$ distributions are shown in the upper panels, while the lower panels present a regional zoom over T\"urkiye (36$^\circ$--42$^\circ$~N, 26$^\circ$--45$^\circ$~E). Positive and negative anomalies indicate enhancements and depletions relative to the quiet-day TEC level, respectively. The color scale represents $\Delta \mathrm{TEC}$ values in TECU.}
\label{fig:fig2}
\end{figure}

Time series plots of $Kp$, $Dst$, and $TEC$ were generated using \texttt{Matplotlib} in Python~3.12.
To illustrate the spatio-temporal evolution of the ionospheric response, two-dimensional $\Delta \mathrm{TEC}$ maps were generated for the pre-storm, main, and recovery phases Figure~\ref{fig:fig2}. The global panels reveal the transition from a relatively quiet ionosphere to a pronounced storm-time depletion, particularly evident across mid-latitude regions during the main phase. The regional zoom over Türkiye highlights that these large-scale structures are still detectable at the$2.5^{\circ}\!\times\!5^{\circ}$ GIM resolution; however, smaller-scale features such as mesoscale travelling ionospheric disturbances or localized storm-enhanced density structures cannot be fully resolved. This emphasizes that the anomalies observed in Figure 2 should be interpreted as broad regional patterns rather than fine-scale ionospheric dynamics. Despite this limitation, the maps clearly demonstrate a coherent depletion signature over Türkiye during the main phase and a partial recovery on the following day, consistent with known mid-latitude storm-time behavior. Cross validation with independent \textit{SWARM} and \textit{FORMOSAT-7/COSMIC-2} electron density (\(N_e\)) data confirmed the regional consistency of storm signatures with previously published results \citep{Remya2025}.

This integrated methodology combines global scale solar wind parameters with regional GNSS derived ionospheric data to characterize storm time responses across mid-latitudes. By aligning multi source datasets in a common temporal framework and applying lag correlation analysis, this study bridges the equatorial mid-latitude coupling perspective emphasized by \citet{Remya2025}. The approach provides a reproducible framework for future monitoring of ionospheric disturbances approaching the expected maximum of Solar Cycle~25.

\section{Results and Discussion}

\begin{figure}[h!]
\centering
\includegraphics[width=0.9\textwidth]{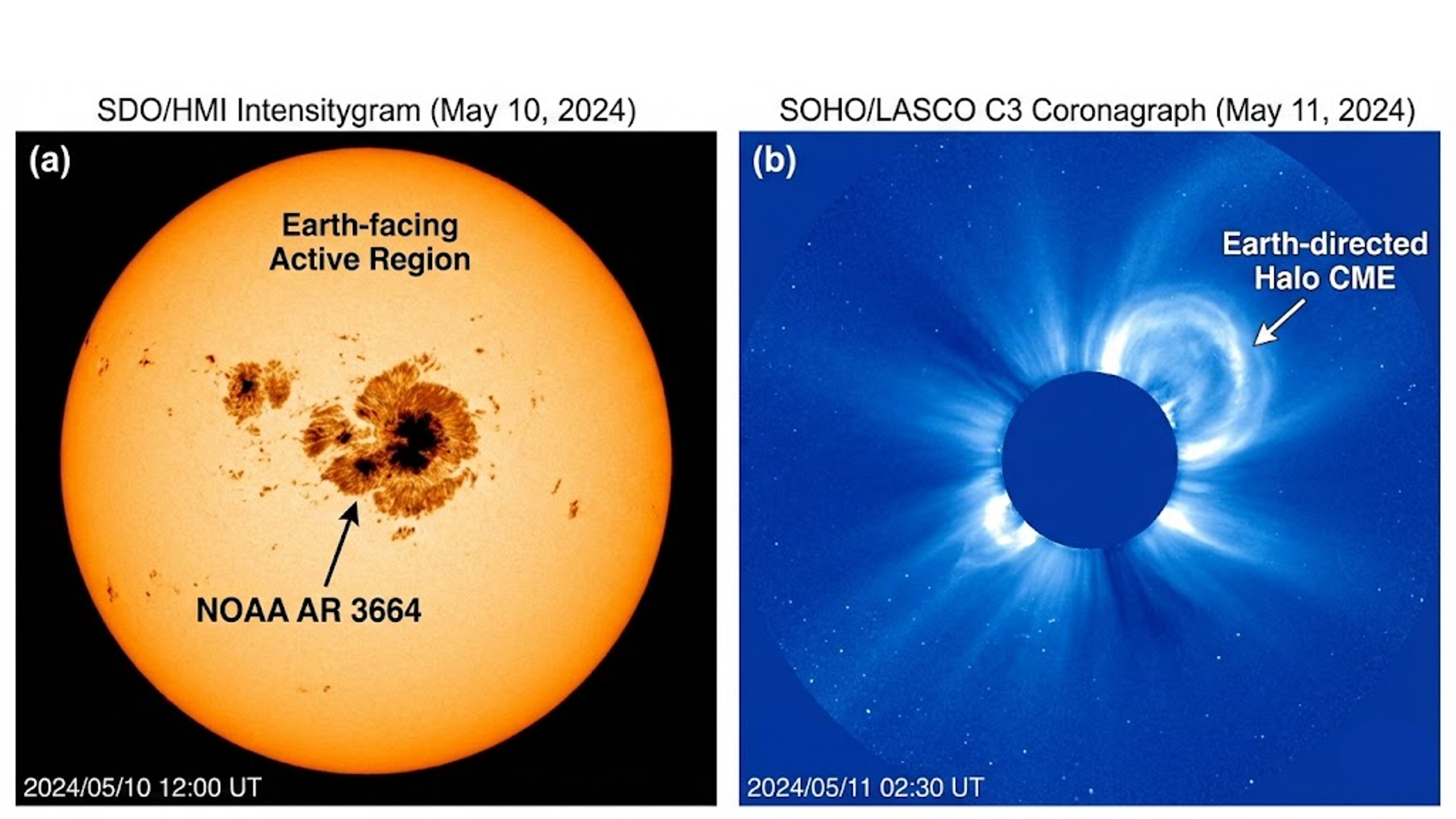}
\caption{(a) SDO/HMI intensitygram showing the Earth-facing active region NOAA AR 3664 on 10 May 2024. (b) SOHO/LASCO C3 image on 11 May 2024 capturing the Earth-directed halo CME associated with AR 3664.}
\label{fig:fig3}
\end{figure}

The geomagnetic storm that occurred between 9–13~May~2024 was among the most geoeffective events of Solar Cycle~25. During this interval, the Interplanetary Magnetic Field (IMF) experienced strong compression and southward turning associated with multiple Earth directed CMEs released from NOAA Active Region 3664 (\citep{Hayakawa2025}, Figure~\ref{fig:fig3}). 

\begin{figure}[h!]
\centering
\includegraphics[width=0.9\textwidth]{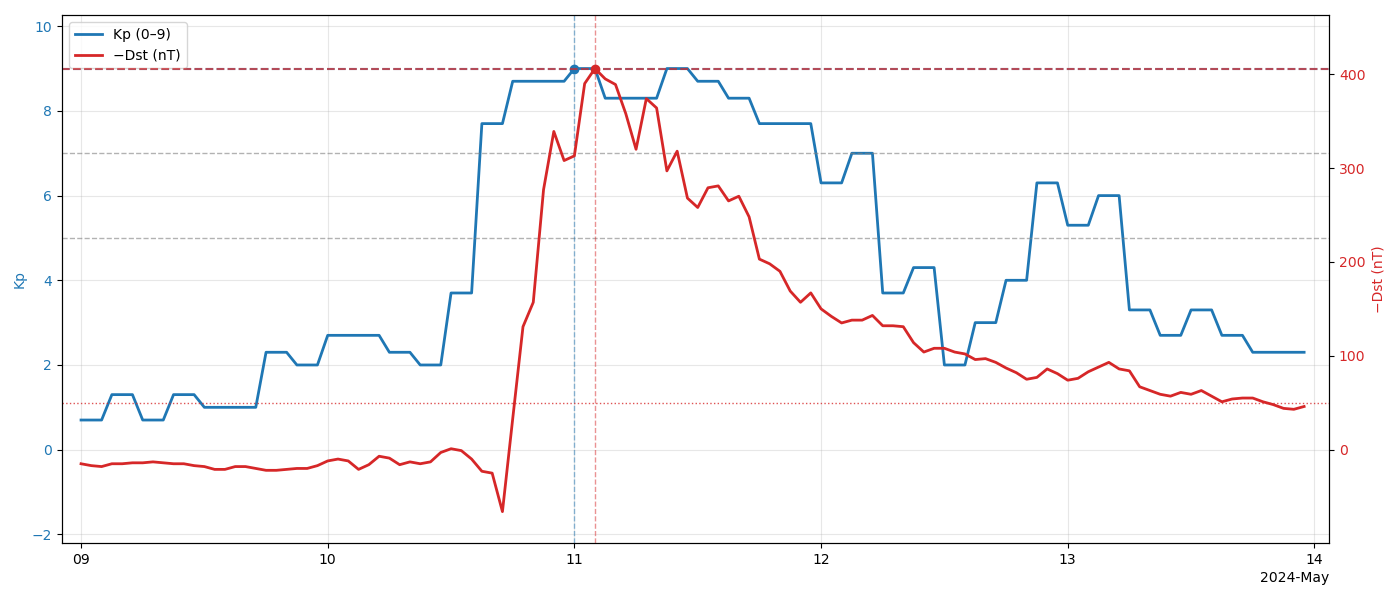}
\caption{
Temporal evolution of geomagnetic indices $Kp$ (blue) and $Dst$ (red) between 9–13~May~2024. A G5-class geomagnetic storm peaked on~May~11, reaching $Kp$~=~9 and $Dst$~=~-400~nT.$Dst$ is plotted as -Dst for clarity.}
\label{fig:kp_dst}
\end{figure}

These successive eruptions drove intense solar wind magnetosphere coupling, culminating in a G5 class geomagnetic storm on May~11, when $Kp$ reached~9 and $Dst$ dropped to approximately~$-400$ nT. The joint evolution of the $Kp$ and $Dst$ indices delineates the storm’s initial, main, and recovery phases, providing a clear temporal context for subsequent ionospheric analyses (Figure~\ref{fig:kp_dst}). For clarity, $Dst$ is plotted as its absolute value (\(-\)Dst) to emphasize magnitude variations.

Figure~\ref{fig:kp_dst} illustrates the temporal evolution of geomagnetic indices $Kp$ (blue) and Dst (red) from 9–13~May~2024. Geophysical conditions in Türkiye’s mid-latitude sector can be placed within this broader context: prior to ~18~UT on 10~May, Kp remained $\leq$3 and Dst hovered near 0~nT, consistent with a quiet ionospheric background. 

Following the sudden storm commencement (SSC) around that time, the storm’s main phase peaked on 11~May when $Dst$ dropped to $\approx$-400~nT and $Kp$ surged to~9 characteristic of a G5 class event. In our GNSS derived TEC data for Türkiye, the main phase corresponded to a sharp reduction in $\Delta$TEC of approximately 20–30\% relative to the 5-day prestorm mean, with the $TEC$ minimum occurring about 2~hours after the Dst minimum. During the recovery phase (after ~00:00~UT on 12~May), $TEC$ values gradually returned to near prestorm levels within the following ~24~hours. Comparative analyses reveal both global consistency and regional variability. For instance, \citet{Lopez2025} reported an atypical TEC depletion over Ecuador (equatorial latitudes) during the same storm, attributed to rapid recombination and plasma instability processes.  

At continental scales, \citet{Danilchuk2025} found significant GNSS positioning errors and enhanced ROTI values extending to mid-latitudes across the Northern Hemisphere, while \citet{Berényi2025} observed pronounced subauroral TEC perturbations over Central and Eastern Europe. Our results strengthen this collective evidence by demonstrating that Türkiye’s mid-latitude ionosphere also underwent strong depletion rather than enhancement, highlighting the latitude dependent and electrodynamically asymmetric nature of the May~2024 superstorm. These findings underscore the importance of incorporating Mediterranean and Anatolian mid-latitude data into global ionospheric storm models, particularly as Solar Cycle~25 approaches its maximum.

\begin{figure}[h!]
\centering
\includegraphics[width=0.95\textwidth]{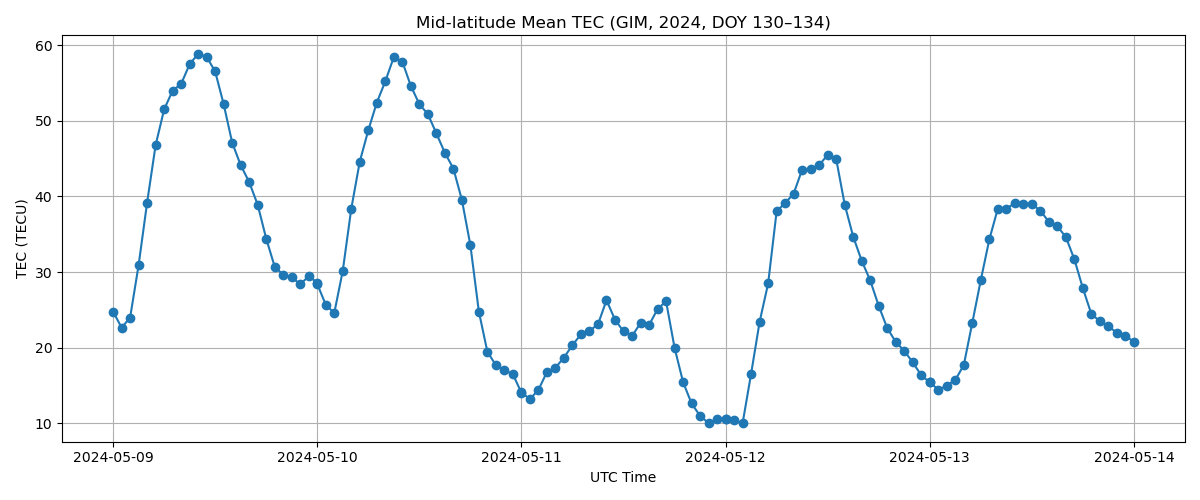}
\caption{
Mid-latitude mean Total Electron Content ($TEC$) derived from Global Ionospheric Maps (GIM) between 9–14~May~2024 (DOY~130–134). A pronounced depletion is observed on 11~May, when TEC values decreased from about 50 to 15~TECU, corresponding to the main phase of the geomagnetic storm shown in Figure~\ref{fig:kp_dst}. The recovery trend after 12~May reflects gradual ionospheric reionization during the storm’s recovery phase.
}
\label{fig:tec_mean}
\end{figure}

The ionospheric response to the May~2024 geomagnetic storm was analysed using Global Ionospheric Maps (GIM) from NASA’s Crustal Dynamics Data Information System (CDDIS), averaged over the region spanning $30^{\circ}$–$60^{\circ}$~N and $25^{\circ}$–$45^{\circ}$~E. Figure~\ref{fig:tec_mean} illustrates the temporal evolution of the mean Total Electron Content (TEC) derived from these GIM data between 9–14~May~2024. Before the storm, TEC values exhibited a regular diurnal pattern, with daytime maxima of about~50~TECU and nighttime minima near~20~TECU, representing typical mid-latitude ionospheric behaviour under quiet geomagnetic conditions.

During the main phase of the storm on 11~May, a pronounced TEC depletion was observed, with values dropping from about 50~TECU to nearly 12–15~TECU (Figure~\ref{fig:kp_dst}). This strong negative phase reflects enhanced ion recombination and suppressed plasma production caused by storm time thermospheric upwelling and composition changes, notably the decrease in the O/N$_2$ ratio \citep{Blagoveshchenskii2013, Remya2025}. The subsequent recovery from 12~May onward marks the gradual re-establishment of plasma density through equatorward transport of ionization and relaxation of disturbed electric fields. 

A minor TEC enhancement on 13–14~May suggests residual electrodynamic activity and partial storm enhanced density (SED) effects, consistent with observations from other mid-latitude regions \citep{Lopez2025, Danilchuk2025}. 
Overall, the $\Delta$TEC amplitude of $\sim$35~TECU confirms this event as one of the most intense mid-latitude ionospheric disturbances of Solar Cycle~25, exhibiting the classical negative storm response followed by a moderate recovery.

\begin{figure}[h!]
\centering
\includegraphics[width=0.95\textwidth]{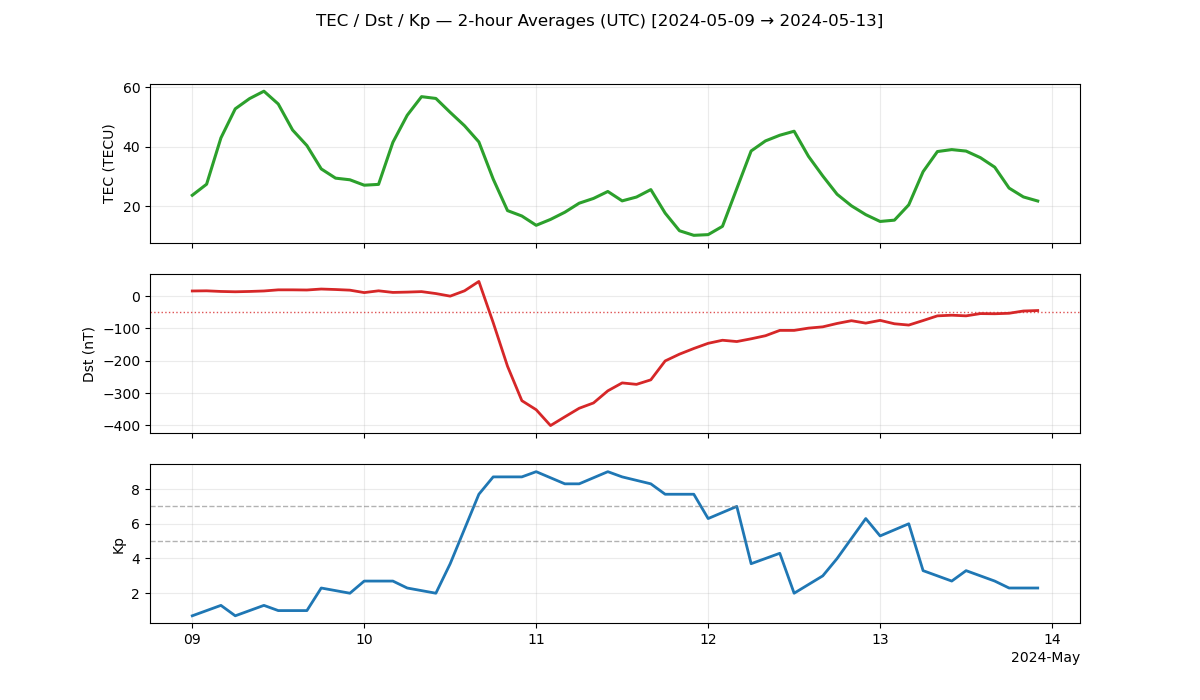}
\caption{
Two hour averaged variations of Total Electron Content (TEC, green), $Dst$ (red), and $Kp$ (blue) indices between 9–13~May~2024. The sharp TEC depletion on 11~May coincides with the peak geomagnetic activity (Dst~$\approx$-400~nT, Kp~$\approx$~9), demonstrating the clear anti correlation between ionospheric electron content and geomagnetic disturbance intensity.}
\label{fig:tec_kp_dst}
\end{figure}

To examine the coupling between geomagnetic and ionospheric activity, two hour averaged variations of Total Electron Content ($TEC$), $Dst$, and $Kp$ indices were analysed for 9–13~May~2024 (Figure~\ref{fig:tec_kp_dst}). The synchronized fluctuations among these parameters reveal a clear temporal relationship between geomagnetic storm phases and ionospheric electron density changes.

To investigate the coupling between geomagnetic and ionospheric activity, two hour averaged variations of Total Electron Content ($TEC$), $Dst$, and $Kp$ indices were analysed for 9–13~May~2024 (Figure~\ref{fig:tec_kp_dst}).  
Before the storm (9–10~May), TEC showed regular day night oscillations between 20–60~TECU, while Kp~$<$~3 and $Dst$ stayed near~0~nT, indicating quiet conditions.  

Around 18:00~UT on~10~May, geomagnetic activity intensified, leading to the main phase on~11~May when Kp reached~9 and Dst fell to~$-400$~nT. Simultaneously, TEC dropped to $\sim$15~TECU, marking a strong negative storm phase driven by CME induced magnetospheric compression and enhanced energy input into the ionosphere thermosphere system \citep{Blagoveshchenskii2013, Lopez2025}.  

After~12~May, $Dst$ and $Kp$ gradually recovered, and TEC showed partial restoration with minor enhancements on~13–14~May, reflecting thermospheric relaxation and ion redistribution by meridional winds \citep{Remya2025}. The synchronized variations among $TEC$, $Dst$, and $Kp$ highlight efficient magnetosphere ionosphere coupling under intense CME forcing, consistent with mid-latitude responses reported by \citet{Danilchuk2025}. Overall, Figure~\ref{fig:tec_kp_dst} reveals a clear anticorrelation between geomagnetic indices and TEC during the May~2024 G5 storm, confirming that mid-latitude ionospheric depletion was primarily governed by enhanced geomagnetic activity.

\begin{figure}[h!]
\centering
\includegraphics[width=0.95\textwidth]{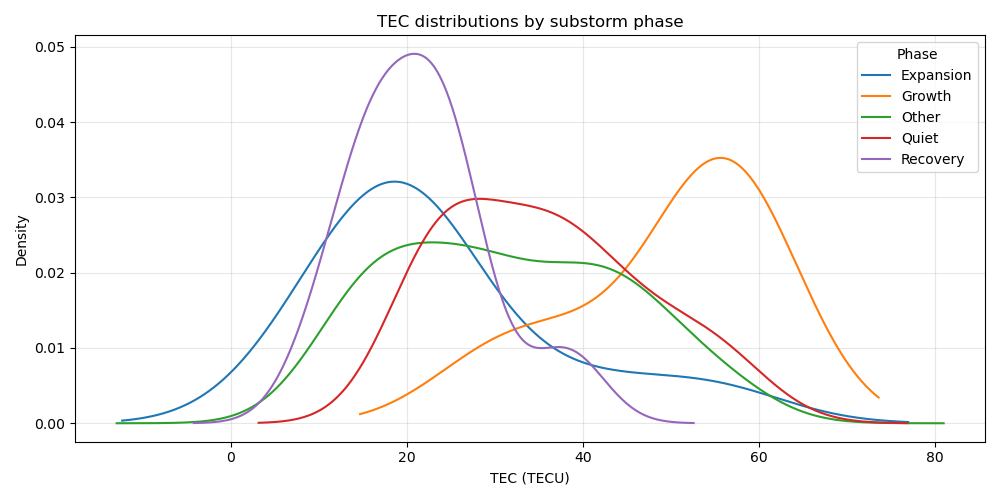}
\caption{
Probability density distributions of Total Electron Content ($TEC$) for different substorm phases during the May~2024 storm. 
The growth phase (orange) shows higher TEC values due to enhanced plasma production, whereas the expansion and recovery phases (blue and purple) shift toward lower TEC, reflecting strong depletion and gradual post-storm relaxation. 
Quiet time conditions (red) display a narrow distribution centered near 35~TECU, representing typical mid-latitude background levels.
}
\label{fig:tec_phase}
\end{figure}

To explore the phase dependent nature of the ionospheric response, the Total Electron Content ($TEC$) data were separated according to the substorm phases defined by geomagnetic indices and solar wind parameters. Figure~\ref{fig:tec_phase} presents the probability density distributions of $TEC$ during the \textit{Growth}, \textit{Expansion}, \textit{Recovery}, and \textit{Quiet} phases of the May~2024 storm. The probability density functions (PDFs) of $TEC$ for different substorm phases reveal a systematic shift toward lower TEC values during the expansion and recovery phases, confirming the dominance of depletion processes under enhanced geomagnetic forcing. Conversely, the growth phase exhibits a broader, high TEC distribution, reflecting increased plasma production before the main disturbance.

Figure~\ref{fig:tec_phase} shows probability density functions (PDFs) of TEC conditioned on substorm phases. The \emph{growth} phase exhibits a broader, high TEC distribution (peaking near 50–60~TECU), consistent with enhanced plasma production prior to the main disturbance. By contrast, the \emph{expansion} and \emph{recovery} phases shift toward markedly lower TEC (modes $\sim$15–25~TECU), indicating storm time depletion and gradual post storm reionization driven by thermospheric upwelling and composition changes (decreased O/N$_2$) and elevated recombination rates \citep{Blagoveshchenskii2013, Remya2025}. Quiet time PDFs are narrow and centered near 30–40~TECU, representing the mid-latitude background over Türkiye. This phase conditioned behaviour aligns with mid-latitude negative storm responses and delayed recoveries reported in regional and hemispheric assessments \citep{Danilchuk2025, Berényi2025}, while contrasting with equatorial case studies where atypical depletion and phase timing reflect distinct electrodynamics \citep{Lopez2025}. Overall, the PDFs corroborate a robust, phase dependent coupling between geomagnetic forcing and ionospheric electron content over Türkiye, reinforcing the need for phase resolved modeling and nowcasting during extreme CME driven storms.

\begin{figure}[h!]
\centering
\includegraphics[width=0.95\textwidth]{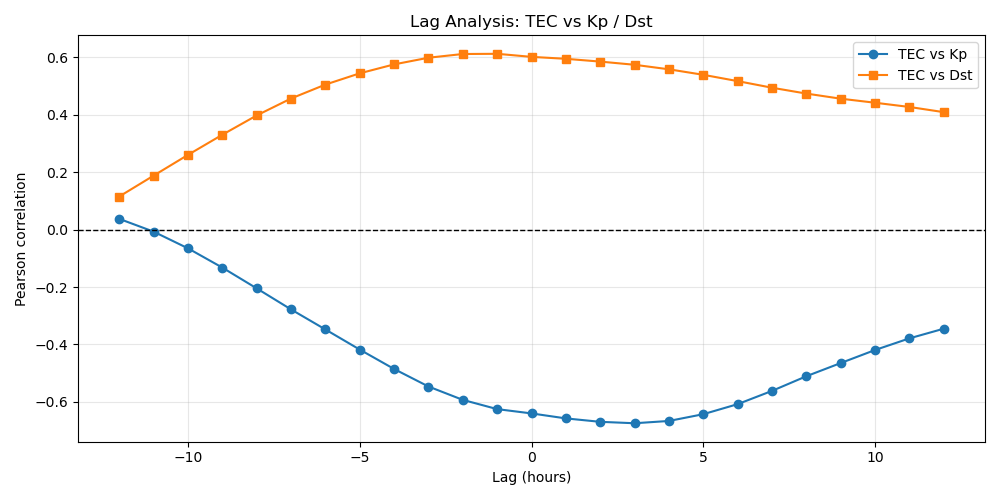}
\caption{
Lag correlation between TEC and geomagnetic indices (9–13~May~2024). Positive lags denote delayed $TEC$ response to geomagnetic activity. The positive correlation with $Dst$ ($r \approx 0.6$ at $+3$~h) indicates a delayed recovery driven by thermospheric processes, while the negative correlation with $Kp$ ($r \approx -0.6$ at $-6$~h) reflects a prompt depletion caused by penetration electric fields.  Together, they demonstrate the two step ionospheric response during the May~2024 G5 storm.}

\label{fig:lag_analysis}
\end{figure}

The temporal relationship between ionospheric Total Electron Content (TEC) and geomagnetic indices was examined using a lag correlation analysis (Figure~\ref{fig:lag_analysis}). This method enables identification of the time delay between magnetospheric forcing and the corresponding ionospheric response during the May~2024 geomagnetic storm.

As shown in Figure~\ref{fig:lag_analysis}, the correlation between $TEC$ and $Dst$ peaks at $r \approx 0.6$ with a $+3$~hour lag, indicating a delayed mid-latitude ionospheric response to geomagnetic forcing. This timing is consistent with energy transfer via thermospheric dynamics and field-aligned currents \citep{Remya2025}. In contrast, the TEC–Kp correlation shows a negative peak ($r \approx -0.6$) around $-6$~hours, suggesting that ionospheric depletion begins shortly before the $Kp$ maximum, driven by prompt penetration electric fields and pre storm electrodynamic adjustments \citep{Lopez2025}.  

The contrasting lag patterns of $Dst$ and $Kp$ highlight the multi scale nature of magnetosphere–ionosphere coupling: $Dst$ traces slower ring current evolution, while $Kp$ captures rapid substorm activity \citep{Blagoveshchenskii2013}. Overall, the results support a two-step response scenario, where initial electric field-driven depletion is followed by delayed thermospheric recovery, consistent with typical mid-latitude storm behavior over Türkiye.

\begin{figure}[h!]
\centering
\includegraphics[width=0.9\textwidth]{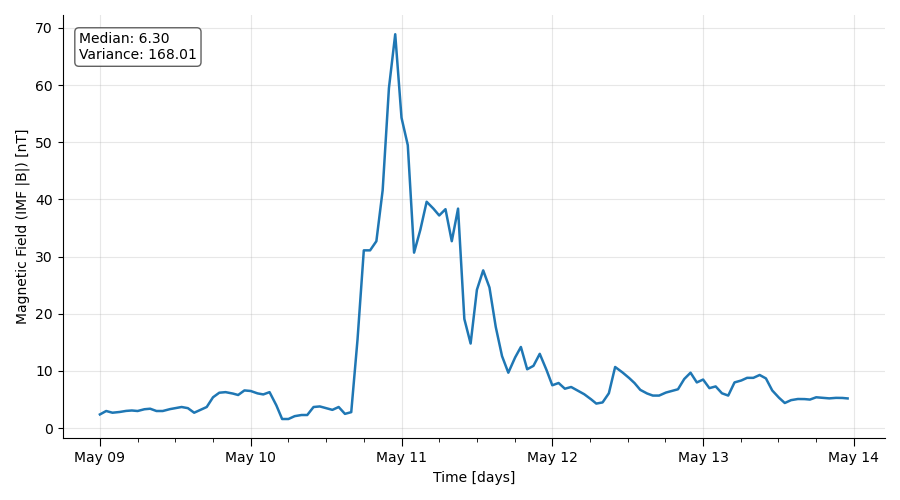}
\caption{
Interplanetary Magnetic Field (IMF) magnitude $|B|$ from OMNI data between 9–14~May~2024.  
A sharp increase to $\sim$70~nT on May~11 marks the arrival of CME driven shocks initiating the geomagnetic storm.  
Median and variance values highlight the strong magnetic compression and subsequent relaxation characteristic of G5 class events.}
\label{fig:mf_intensity}
\end{figure}

The interplanetary magnetic field (IMF) strength, obtained from the OMNI database with one hour temporal resolution, provides essential context for evaluating the geomagnetic conditions preceding and during the May 2024 event (Figure~\ref{fig:mf_intensity}).  
Magnetic field magnitude ($|B|$) was analysed to trace the arrival and propagation of coronal mass ejections (CMEs) responsible for driving the storm.

Figure~\ref{fig:mf_intensity} shows the temporal variation of the IMF magnitude ($|B|$) between 9–14~May~2024.   Prior to 10~May, $|B|$ remained near 5~nT under quiet solar-wind conditions. A sharp increase to $\sim$70~nT around 21:00~UT on~May~10 marks the arrival of a fast Earth-directed CME shock \citep{Hayakawa2025}, coinciding with the onset of intense geomagnetic activity (see Figure~\ref{fig:kp_dst}). The elevated variance (168~nT$^2$) indicates strong magnetic turbulence within the CME sheath, consistent with multi component ejecta reported for this event \citep{Remya2025, Lopez2025}. Following the peak, $|B|$ gradually decreased to 10–15~nT as the interplanetary field relaxed during recovery. The IMF evolution mirrors the $Dst$ and $Kp$ trends, confirming that the May~2024 superstorm was driven by compound CME interactions. This pronounced magnetic compression provided the primary energy source for the magnetosphere–ionosphere coupling that produced the observed mid-latitude TEC depletions.

\begin{figure}[h!]
\centering
\includegraphics[width=0.95\textwidth]{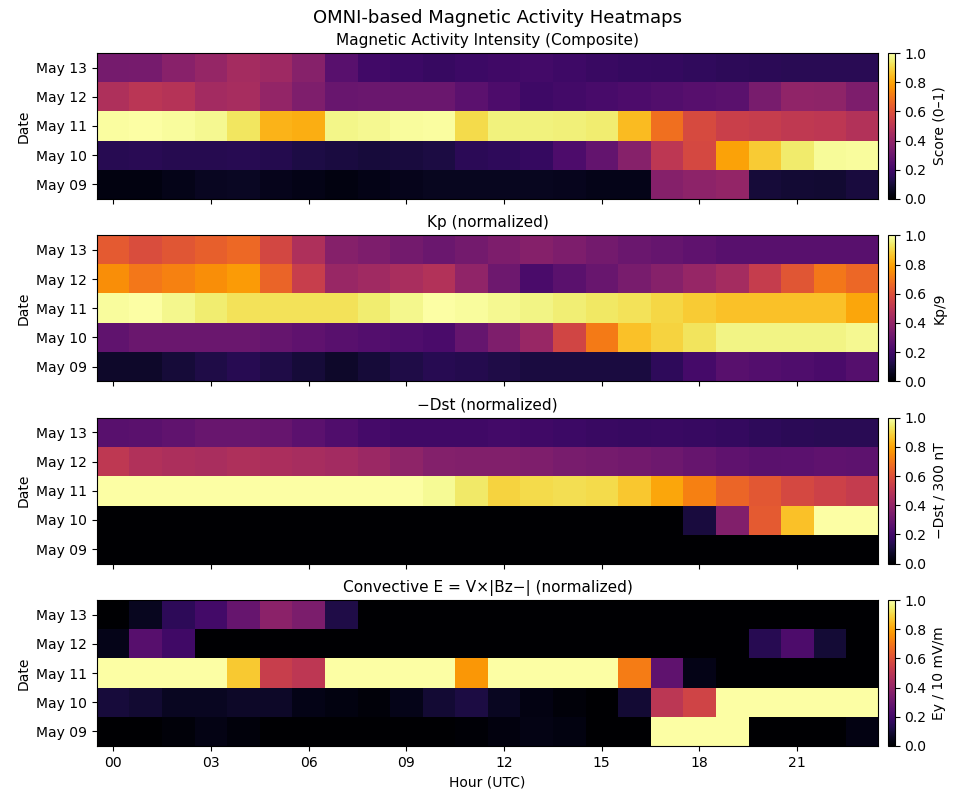}
\caption{OMNI based magnetic activity heatmaps showing composite storm intensity, Kp, Dst, and convective electric field ($E_y = V \times |B_z|$). Enhanced activity on May 11 marks the main geomagnetic disturbance associated with the CME shock arrival.}
\label{fig:omni_heatmaps}
\end{figure}

To better visualize the temporal evolution of geomagnetic activity during the May 2024 storm, a series of heatmaps were generated from OMNI one hour data sets (Figure~\ref{fig:omni_heatmaps}). These maps display normalized variations of Kp, Dst, and the convective electric field ($E_y = V \times |B_z|$), providing a comprehensive view of the magnetospheric forcing mechanisms over the 9–13 May 2024 period.

Figure~\ref{fig:omni_heatmaps} presents OMNI based heatmaps of magnetic activity indices during 9–13~May~2024. The composite panel highlights a sharp enhancement on~May~11, coinciding with the storm’s main phase. Normalized Kp and $-|Dst|$ show nearly synchronous peaks that persist into~May~12–13, indicating sustained magnetospheric convection and multi-step CME interactions \citep{Lopez2025, Hayakawa2025}. The convective electric field ($E_y = V \times |B_z|$) exhibits short but intense bursts ($>10$ mV/m), corresponding to southward IMF and high solar wind speed conditions \citep{Remya2025}. The alignment of these $E_y$ enhancements with Kp and Dst peaks confirms efficient solar wind magnetosphere coupling. Overall, the heatmaps reveal a classic three stage storm sequence quiet preconditioning, intense main phase, and gradual recovery consistent with empirical storm models \citep{Blagoveshchenskii2013}. The timing of peak $E_y$ and magnetic activity aligns closely with the onset of mid-latitude TEC depletion discussed in Figures~\ref{fig:tec_mean}–\ref{fig:tec_phase}.

\section{Conclusions}

The mid-latitude ionospheric response to the 11~May~2024 geomagnetic superstorm over Türkiye exhibited a clear and intense depletion, with TEC values decreasing from approximately 50~TECU to 15~TECU, followed by a gradual recovery. When compared with the equatorial observations reported by \citet{Lopez2025} over the Galápagos, both studies confirm that the storm produced sharp TEC depletions correlated with strong geomagnetic disturbances ($Kp$ and $Dst$). However, while the equatorial sector showed a negative storm effect associated with the weakening of the Equatorial Ionization Anomaly (EIA) and Electrojet (EEJ), the mid-latitude response observed here was more regular yet stronger in magnitude.

This contrast underscores the latitude dependent nature of storm time ionospheric dynamics: equatorial regions are dominated by complex electrodynamic processes that can yield counterintuitive behaviours, whereas mid-latitude regions exhibit a more direct and depletion driven response linked to geomagnetic forcing. Our results therefore, provide new insight into storm time ionospheric behaviour over Türkiye, a region often underrepresented in global ionospheric studies.

These findings emphasize the importance of multi latitude and phase resolved analyses for improving empirical and physics based models of ionospheric variability during extreme solar events. Such integrative approaches are essential for enhancing the reliability of space weather forecasting and mitigating its impacts on GNSS and communication systems.

\section*{Acknowledgements}
The authors would like to thank the anonymous reviewers for their constructive comments and valuable suggestions that significantly improved the quality of this manuscript. Furthermore, the authors acknowledge the use of data products from NASA's OMNIWeb and CDDIS archives. Interplanetary and geomagnetic parameters were obtained through the OMNI dataset provided by the NASA/GSFC Space Physics Data Facility (SPDF). Global Ionospheric Maps (GIMs) in IONEX format were accessed via the NASA CDDIS repository. We also thank the developers and data providers of these open-access resources for their continuous efforts in maintaining and distributing high-quality geospace datasets. This study was conducted using publicly available data, and no proprietary data were employed.

%%%%%%%%%%%%%%%%%%%%%%%%%%%%%%%%%%%%%%%%%%%%%%%%%%%%%%%%%%%%%%%%%%%%%%%%%%%%%

\bibliography{References}

\end{document}